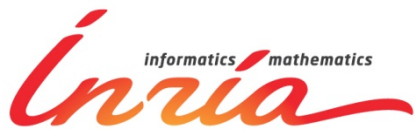

# Evaluation of Synchronous and Asynchronous Reactive Distributed Congestion Control Algorithms for the ITS G5 Vehicular Systems

Oyunchimeg Shagdar

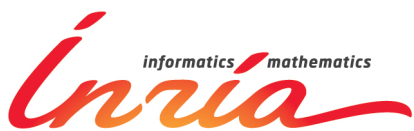

# Evaluation of Synchronous and Asynchronous Reactive Distributed Congestion Control Algorithms for the ITS G5 Vehicular Systems

Oyunchimeg Shagdar[1]

**Abstract:** The IEEE 802.11p is the technology dedicated to vehicular communications to support road safety, efficiency, and comfort applications. A large number of research activities have been carried out to study the characteristics of the IEEE 802.11p. The key weakness of the IEEE 802.11p is the channel congestion issue, where the wireless channel gets saturated when the road density increases. The European Telecommunications Standardization Institute (ETSI) is in the progress of studying the channel congestion problem and proposed so-called Reactive Distributed Congestion Control (DCC) algorithm as a solution to the congestion issue. In this report we investigate the impacts of the Reactive DCC mechanism in comparison to the conventional IEEE 802.11p with no congestion control. Our study shows that the Reactive DCC scheme creates oscillation on channel load that consequently degrades communication performance. The results reveal that the channel load oscillation is due to the fact that in the Reactive DCC, the individual CAM (Cooperative Awareness Message) controllers react to the channel congestion in a synchronized manner. To reduce the oscillation, in this report we propose a simple extension to Reactive DCC, Asynchronous Reactive DCC, in which the individual CAM controllers adopt randomized rate setting, which can significantly reduce the oscillation and improve the network performance.

[1] Researcher, Project-Team RITS, Inria Paris-Rocquencourt – Oyunchimeg.Shagdar@inria.fr

# Evaluation of Synchronous and Asynchronous Reactive Distributed Congestion Control Algorithms for the ITS G5 Vehicular Systems

**Résumé :** L'IEEE 802.11p est la technologie dédiée à la communication des véhicules pour soutenir la sécurité routière, l'efficacité et les applications de confort. Un grand nombre d'activités de recherche ont été menées pour étudier les caractéristiques de l'IEEE 802.11p. La principale faiblesse de l'IEEE 802.11p est la congestion de canal, où le canal se sature lorsque la densité de la route augmente. L'Institut européen de normalisation des télécommunications (ETSI) est en train d'étudier le problème et proposer l'algorithme « Réactif Congestion Control Distribué (DCC) » comme une solution. Dans ce rapport, nous étudions les effets du mécanisme Réactive DCC par rapport à l'IEEE 802.11p classique sans contrôle de congestion. Notre étude montre que Réactif DCC génère une oscillation de la charge de canal qui se dégrade par conséquent les performances de la communication. Les résultats révèlent que l'oscillation de la charge de canal est dû au fait que, dans le Réactif DCC, les individuels contrôleurs du CAM (Cooperative Awareness Message) réagissent à la congestion de canal d'une manière synchronisée. Pour réduire l'oscillation, dans ce rapport, nous proposons une extension simple du Réactif DCC, Asynchrone Réactif DCC, dans lequel les individuels contrôleurs de CAM adoptent réglage de la fréquence aléatoire, ce qui peut réduire de manière significative l'oscillation et d'améliorer la performance du réseau.

**Mots clés :** IEEE 802.11p, la congestion de canal, congestion control distribué, simulation

# 1. Introduction

Wireless communication is expected to play an important role for road safety, efficiency, and comfort of road users [1]. To support such ITS applications the IEEE 802.11p [2] (ETSI ITS G5 [3] for the European usage) is standardized for V2X communications using the 5.9 GHz frequency bands. The IEEE 802.11p has been the focus of a great number of R&D activities, and its applicability to road safety and efficiency applications have been tested in some projects. The key weakness of the IEEE 802.11p is the channel congestion problem, where channel is saturated when the number of the 802.11p-equipped vehicles is large. This problem is obviously due to the limited resource at the 5.9 GHz band, but also because all the vehicles are expected to periodically broadcast CAMs, which are needed for collision avoidance but tend to load the wireless channel.

A number of DCC algorithms such as Reactive DCC [4], LIMERIC [5] and AIMD adaptive control [6] have been proposed. The key differences lie in their ways of controlling the communication parameters. Having the CAM generation rate as the control parameter, the reactive DCC controls the rate following a parameter table; LIMERIC controls the rate based on a linear adaptive algorithm, and AIMD algorithm control the rate in a similar manner as TCP.

On the other hand, channel busyness ratio (CBR), which is the ratio of the time the channel perceived as busy to the monitoring interval, is the commonly agreed metric used to characterize channel load. Since the wireless channel is shared by the ITS-S that are in the vicinity of each other, CBR monitored at such ITS-Ss take similar values. As a consequence, the ITS-Ss may take synchronized reactions to the channel load, e.g., the ITS-Ss reduce/increase the transmission rate at around the same time. The first contribution of this work is thus to study such a synchronized DCC behavior observed in reactive DCC algorithm. We pay an attention on the following different possible reactions of the CAM generator, which is responsible for adjusting the message generation rate as a means of DCC.

Timer handling: In general, a transmission of a CAM is triggered by a timer, which is set to the CAM interval. Hence, upon being informed with a new CBR value (at an arbitrary point of time), the CAM generator may i) wait the expiration of the on-going timer and set the timer to the new CAM interval or ii) cancel the on-going timer and set it to the new CAM interval. We call the former and latter behaviors as Wait-and-Go and Cancel-and-Go.

Interval setting: As mentioned above, CBR measured for the shared channel may lead to the situation where the nearby ITS-Ss increase/decrease the CAM interval at around the same time. This is especially true for the reactive DCC algorithm, which controls the rate following a table. Therefore, one can think of avoiding such a synchronized behavior by applying random intervals. Hence, we can imagine 2 possible behaviors: upon reception of a new CBR value, the CAM generator sets the message generation interval to i) the value (say new_CAM_interval) provided by the table or ii) a random value (e.g., taken from the range [0, new_CAM_interval]) for the first packet and then follows the table. We call the former and latter behaviors as Synchronized and Unsynchronized.

Considering the above-mentioned behaviors of the CAM generator, we obtain the following 4 different versions of Reactive DCC:

- DccReactive-1: Wait-and-Go & Synchronized
- DccReactive-2: Cancel-and-Go & Synchronized
- DccReactive-3: Wait-and-Go & Unsynchronized
- DccReactive-4: Cancel-and-Go & Unsynchronized

The first contribution of this work is hence, to study and compare the performances of these different versions of reactive DCC to understand the synchronization issue and the underlying reasons.

Second contribution of this work is to have a close look to channel load characterization. While it is commonly agreed that CBR should be monitored over a certain interval (e.g., 100 ms), it is not clear if channel load should be characterized with the current value of CBR or it should also consider the past CBR values. To study this aspect, we define channel load (CL) as follows.

$$CL_n = (1 - \alpha) \times CL_{n-1} + \alpha \times CBR_n \tag{1}$$

Here, $CBR_n$ is CBR measured at the nth monitoring interval, $CL_n$ is the channel load calculated upon measurement of $CBR_n$. As can be seen in (1), the weight factor, α, plays the key role for defining whether the channel load should consider only the last CBR or should pay an attention on its history. Obviously by choosing α=1, channel load is characterized by the "current" channel condition. In our study, we evaluate the performances of reactive DCC for different values of α.

The third contribution of this work is to study the DCC performance in road systems, which consist of ITS-Ss with different levels of sensing capability. Specifically, we consider that ITS-Ss sense the wireless channel at different levels, and as consequence, they perceive CL differently and react differently. To realize this study, ITS-Ss in the simulations are provided with random sensitivity offset values in the range of [-6, +6] dBm.

To summarize, the contributions of this work are as follows

Contribution 1: Study on synchronization issue of DCC control.

Contribution 2: Study on channel load characterization.

Contribution 3: Study on non-identical sensing capabilities.

## 2. Used Simulation Tools

The work is carried out using the open discrete event simulation environment NS3 (version 3.21) [7], and the traffic simulator SUMO (version 0.22) [8]. The key simulation modules, which are relevant to this work, are illustrated in FIGURE 1, where the modules written in red are newly developed software.

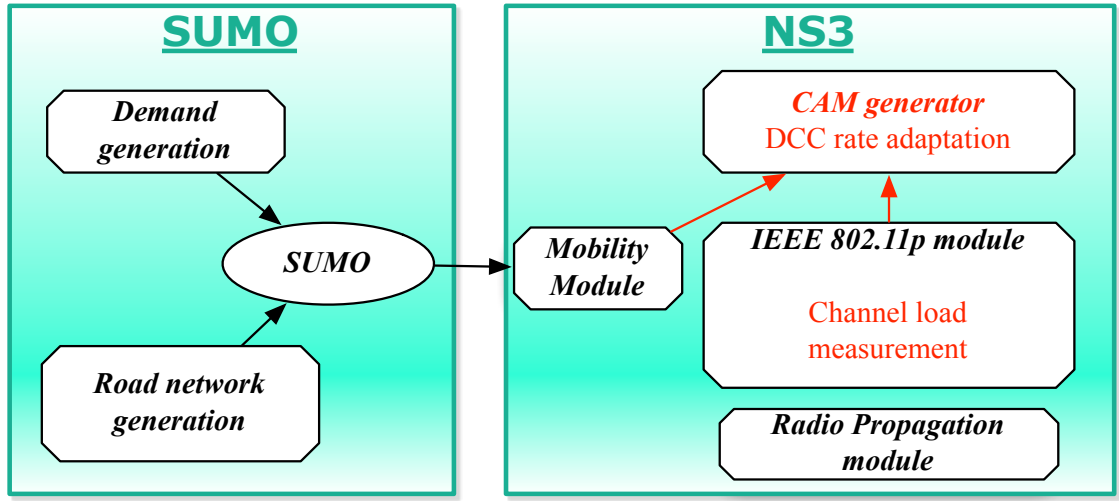

FIGURE 1: Simulators and the key modules relevant to the work.

## 2.1 NS3

The latest stable version of NS3 at the time of writing, NS-3.21, is used in this work. Among a number of new functionalities, it includes the WAVE system, which has the IEEE 802.11p (ITS-G5). The system follows the TCP/IP communication architecture. The key software components used in our simulations are a CAM generator, UDP/IP, IEEE 802.11p, radio propagation, and mobility modules.

The CAM generator is a newly developed module, which takes position and mobility information from the mobility module and periodically generates CAMs. The module is implemented with DCC rate adaptation algorithms. This work focuses on the reactive DCC algorithm. When the reactive DCC module is provided with a CL value (see (1)), it adjusts the CAM generation interval following the parameter table.

Messages generated at the CAM generator processed by the UDP and IP modules, and received at the IEEE 802.11p MAC. While, BTP/ GeoNetworking protocols are standardized in ETSI, utilizing UDP/IP is equivalent to utilizing BTP/Geonetworking for the objective of studying channel congestion caused by 1-hop broadcast messages (CAM). It should be noted that since the header lengths of UDP/IP and BTP/Geonetworking are different, the necessary message length adjustment is made at the CAM generator such that the length of the frames transmitted on the wireless channel have the same length to the case using BTP/Geonetworking.

The PHY layer of NS3 is extended with a CBR monitoring functionality, which monitors the channel activities and calculates CL. Since NS3 is an event-based simulator, the CBR monitoring module exploits the event notifications installed in NS3. In addition, the module holds a timer and calculates CBR in every Tmonitor interval following (2). It should be mentioned that timer setting is made independently at each ITS-S, and hence the CL notifications to the CAM generator is not synchronized among the individual ITS-Ss.

$$CBR = \frac{\sum T_{busy}}{T_{monitor}} \qquad (2)$$

NS3 mobility module is responsible for mobility of ITS-Ss, and it is the interface of NS3 with the SUMO traffic simulator.

## 2.2 SUMO

The SUMO traffic simulator is used to generate road network and traffic following user-specified scenarios. The outputs of the traffic simulator are converted in a file format readable by the mobility module of the NS3 simulator.

## 3. Used Simulation Tools

Unless otherwise noted, the communication and road parameters take the values listed in Table 1

**Table 1. Simulation Parameters.**

| Parameters | Value |
|---|---|
| Communication | |
| CAM default Tx rate | 10 Hz |
| CAM message size | 400 Bytes |
| Tx Power | 23 dBm |
| ED$^{threshold}$ | -95 dBm |
| EDCA Queue / TC | 1 DENM / 3CAM |
| Modulation scheme | QPSK ½ 6 Mbps |
| Antenna pattern | Omnidirectional, gain = 1dBi |
| Access technology | ITS G5A |
| ITS G5 Channel | CCA |
| Fading model | LogDistance, exponent 2 |
| Road network | |
| Lane width | 3 m |
| Lanes in-flow | 3 |
| Lanes contra-flow | 3 |
| DCC parameters | |
| CBR monitor interval (T$_{monitor}$) | 100 ms |
| (see (1)) | 1 |

The parameter table of the reactive DCC algorithm is shown in Table 2.

**Table 2. DCC Reactive Parameters.**

| States | CL(%) | $T_{off}$ |
|---|---|---|
| Relaxed | 0%≤CL<19% | 60 |
| Active_1 | 19%≤CL<27% | 100 |
| Active_2 | 27%≤CL<35% | 180 |
| Active_3 | 35%≤CL<43% | 260 |
| Active_4 | 43%≤CL<51% | 340 |
| Active_5 | 51%≤CL<59% | 420 |
| Restricted | CLR59% | 460 |

### 3.1 Simulation Scenarios

The work is carried out for homogenous highway scenarios. Table 3 lists the road configuration. As shown in the table and illustrated in FIGURE 2: Illustration of a homogenous highway scenario. RSUs are installed in every 100 m on the middle lane.

The scenario consists of sparse, medium, dense, and extreme density classes; the density parameters are listed in Table 4. Density parameter for homogenous highway scenario..

**Table 3. Road Configuration.**

| Class | Inter-vehicle distance |
|---|---|
| Highway length | 1000 m |
| Lanes/Directions | 3 lanes / 2 directions |
| RSU inter-location | 100 m |
| Vehicle size | 2m x 5m |

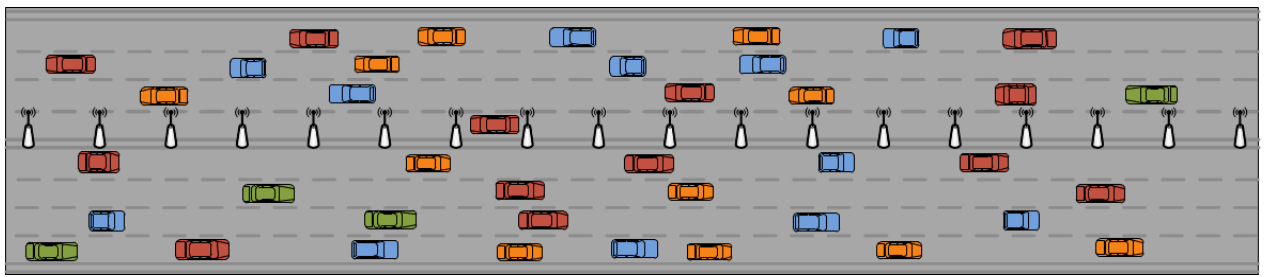

FIGURE 2: Illustration of a homogenous highway scenario.

**Table 4. Density parameter for homogenous highway scenario.**

| Class | Inter-Vehicle distance | Mobili |
|---|---|---|
| Sparse | 100 m inter-distance (3 lanes/ 2 directions) | Static/Mobile |
| Medium | 45 m inter-distance (3 lanes / 2directions) | Static/Mobile |
| Dense | 20 m inter-distance (3 lanes / 2directions) | Static/Mobile |
| Extreme | 10 m inter-distance (3 lanes / 2directions) | Static |

### 3.2 Performance Metrics

As described in Section 1 this work makes three contributions 1) study on synchronization issue of DCC control, 2) study on channel load characterization, and 3) study on non-identical sensing capabilities. Following metrics are used for performance investigations.

- Packet delivery ratio (PDR): the ratio of the number of received packets over the number of transmitted (generated) packets. PDR is measured at individual ITS-S (vehicles and RSUs) targeting CAMs transmitted by each mobile ITS-S (i.e., vehicles).
- Packet Inter-Reception time (PIR): time gap between consecutive CAM messages. PIR is measured at individual ITS-Ss for received CAMs from each mobile ITS-S.
- Number of transmissions: the total number of CAM transmissions is counted for 20 milliseconds of time bins.
- CBR: the average CBR is calculated for 20 milliseconds of time bins.
- Jain's fairness index is calculated for the total number of transmissions from individual mobile ITS-Ss.

## 4. Study on DCC Synchronization Issue

This section evaluates the performances of the four different versions of Reactive DCC: DccReactive-1, DccReactive-2, DccReactive-3, and DccReactive-4 for homogeneous static highway scenario. The performances of these mechanisms are compared against that of DccOff, which is the ITS-G5 system without distributed congestion control.

### 4.1 Evaluation of Packet Delivery Ratio

FIGURE 3-FIGURE 6 plot the average packet delivery ratio (PDR) of the reactive DCC mechanisms in contrast to that of DccOff. The horizontal axis is the distance between the receivers and the transmitters. First of all, we note that DccOff shows an excellent PDR performance in the sparse scenario (see FIGURE 3), where obviously the channel is not congested. The channel congestion becomes an issue for medium, dense and extreme density classes, where PDR degrades down to 10% in DccOff. DccReactive mechanisms show better PDR than DccOff. The PDR improvement is much more significant for unsynchronized DCC schemes (DccReactive-3 and -4) than for synchronized scheme (DccReactive-1 and -2). For timer handling, Cancel-and-Go schemes show poorer performances (DccReactive-2 in comparison to DccReactive-1 and DccReactive-4 in comparison to DccReactive-3).

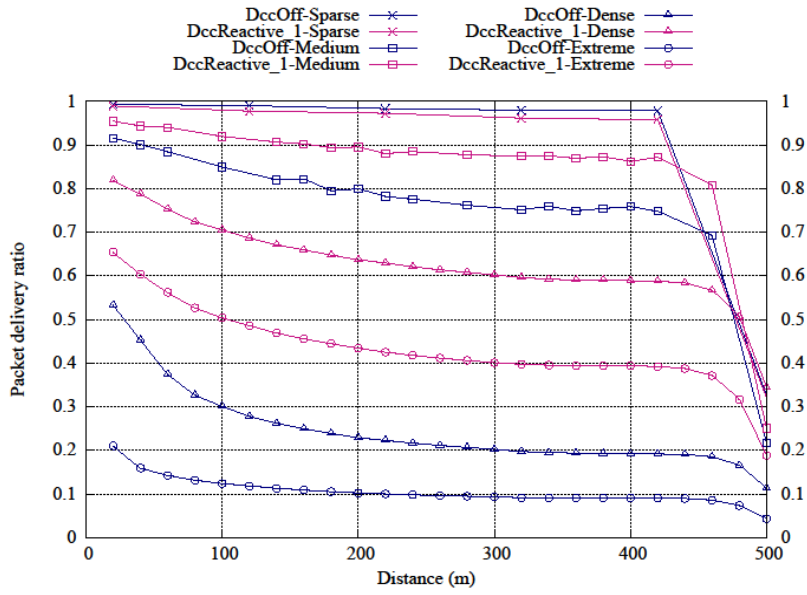

FIGURE 3: Comparison of PDR performances of DccReactive-1 and DCC-Off.

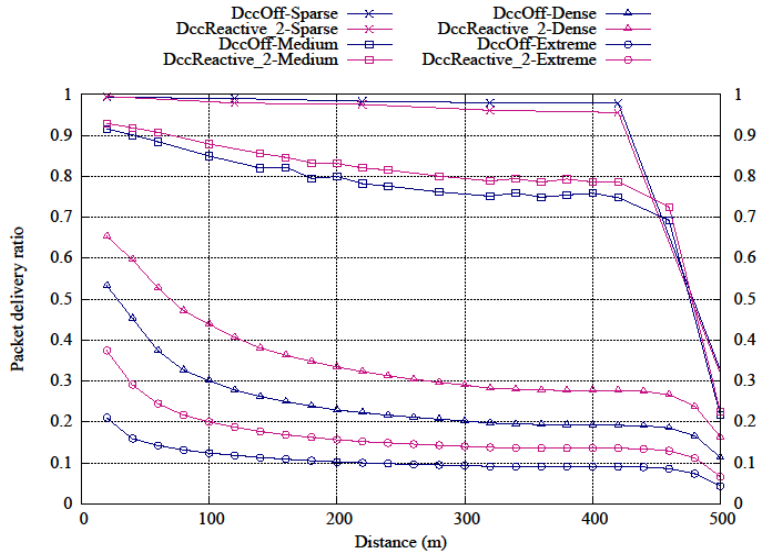

FIGURE 4: Comparison of PDR performances of DccReactive-2 and DCC-Off.

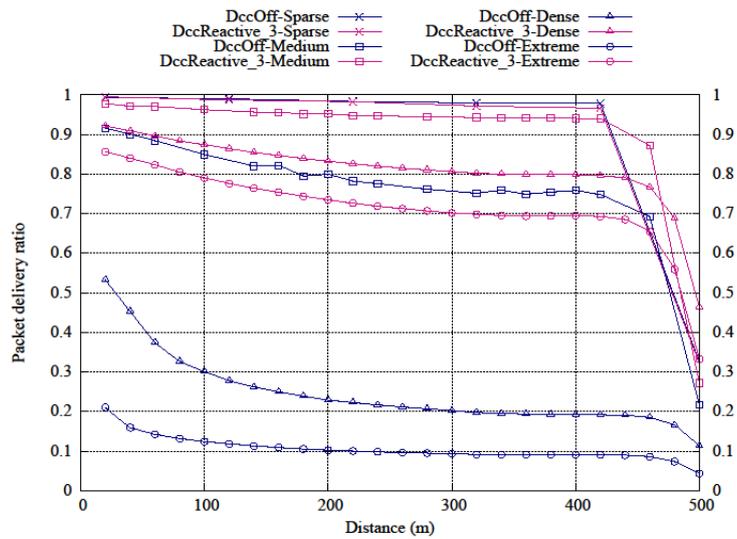

FIGURE 5: Comparison of PDR performances of DccReactive-3 and DCC-Off.

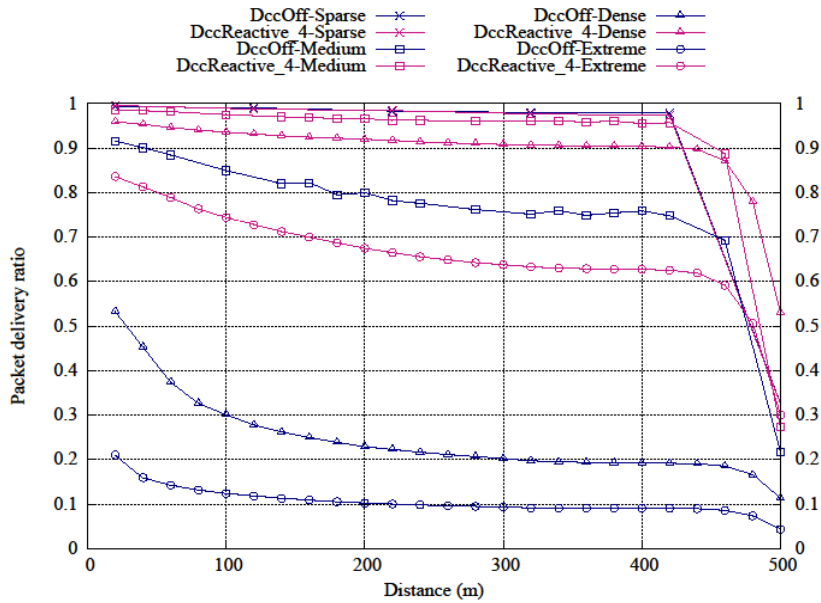

FIGURE 6: Comparison of PDR performances of DccReactive-4 and DCC-Off.

## 4.2 Evaluation of Packet Inter-Reception Time

FIGURE 7 - FIGURE 10 plot the average packet inter-reception time (PIR) of the reactive DCC mechanisms in contrast to that of DccOff. Similar to the PDR case, DccOff shows an excellent PIR performance in the sparse scenario, but the performance largely degrades for higher density classes and it can exceed 1 second in the extreme density class. The reactive DCC mechanisms show better or worse

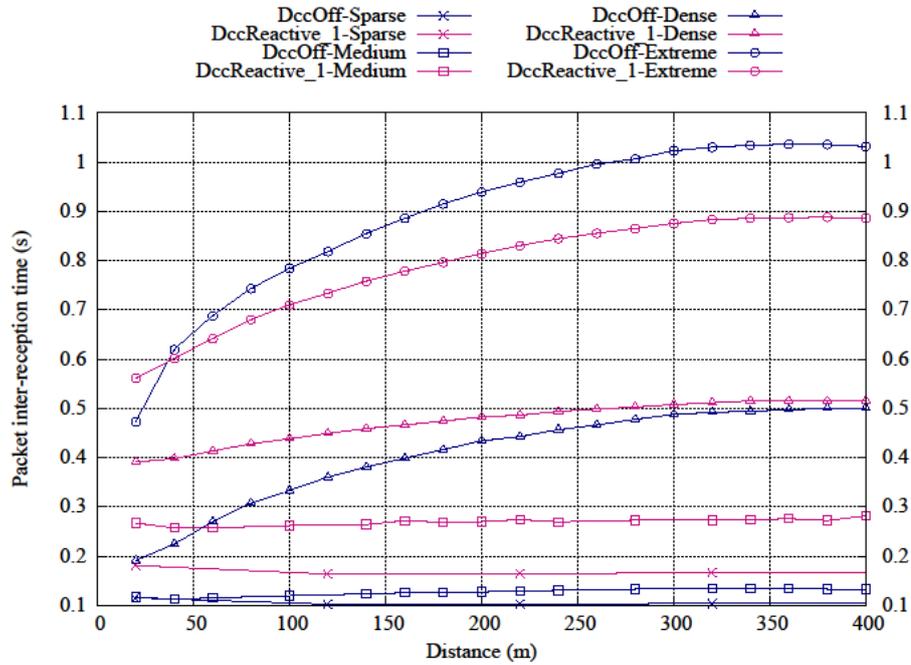

FIGURE 7: Comparison of PIR performances of DccReactive-1 and DCC-Off.

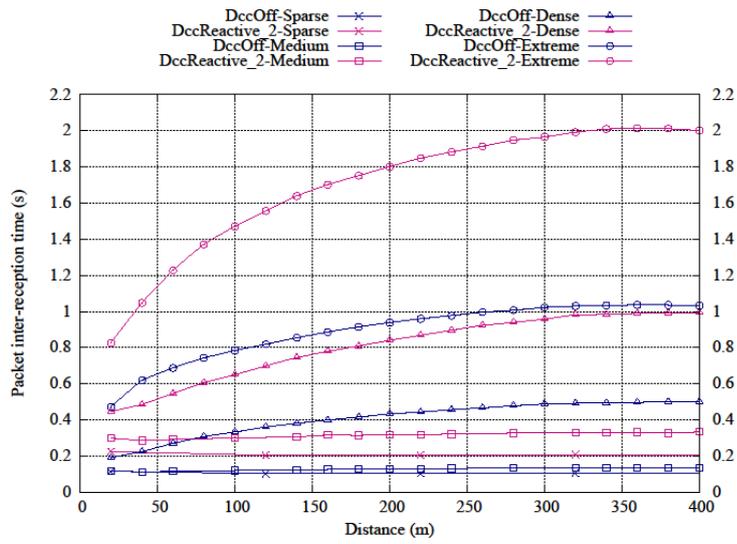

FIGURE 8: Comparison of PIR performances of DccReactive-2 and DCC-Off.

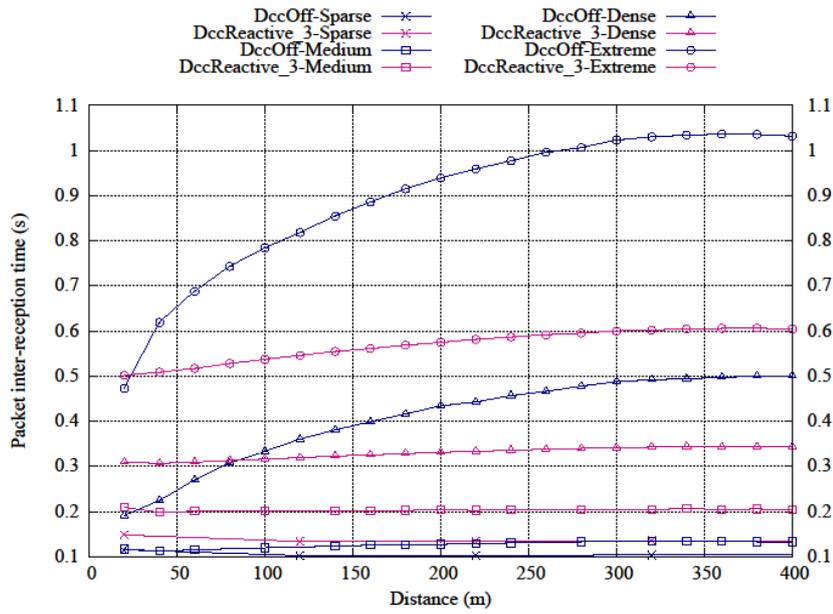

FIGURE 9: Comparison of PIR performances of DccReactive-3 and DCC-Off.

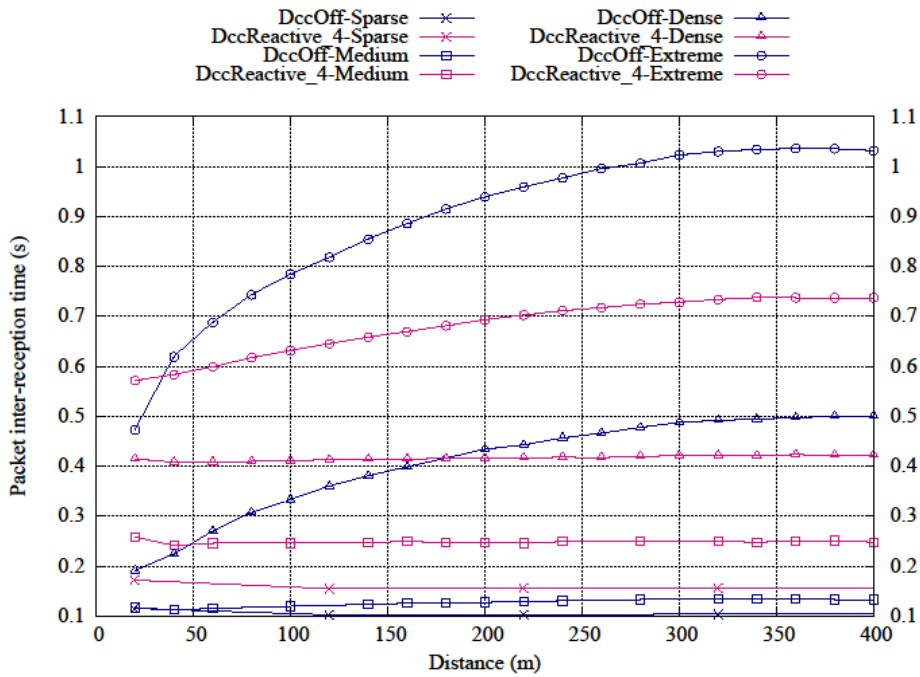

FIGURE 10: Comparison of PIR performances of DccReactive-4 and DCC-Off.

### 4.3 Channel Load Evaluation

FIGURE 11- FIGURE 14 plot the total number of transmissions during a 5-seconds time interval for the dense scenario. Similarly, FIGURE 15-FIGURE 18 plot the channel load measured during a 5-seconds time interval for the dense scenario. Roughly speaking, the number transmissions during 20-milliseconds of time bins takes values in the range of [27, 35] for DccOff. In contrast, the value oscillates in the range of [5, 30], [0, 35], [10, 20] and [7, 12] for DccReactive-1, -2, -3, and -4, mechanisms respectively. Similar oscillated behaviors can be observed for the measured CBR (FIGURE 15-FIGURE 18). Specifically, in the dense scenario, CBR of DccOff is stable at 0.84%. In contrast, the CBR value oscillates in the range of [0.2, 0.8], [0.1, 0.7], [0.55, 0.8], and [0.4, 0.6] for DccReactive-1, -2, -3, and -4, respectively.

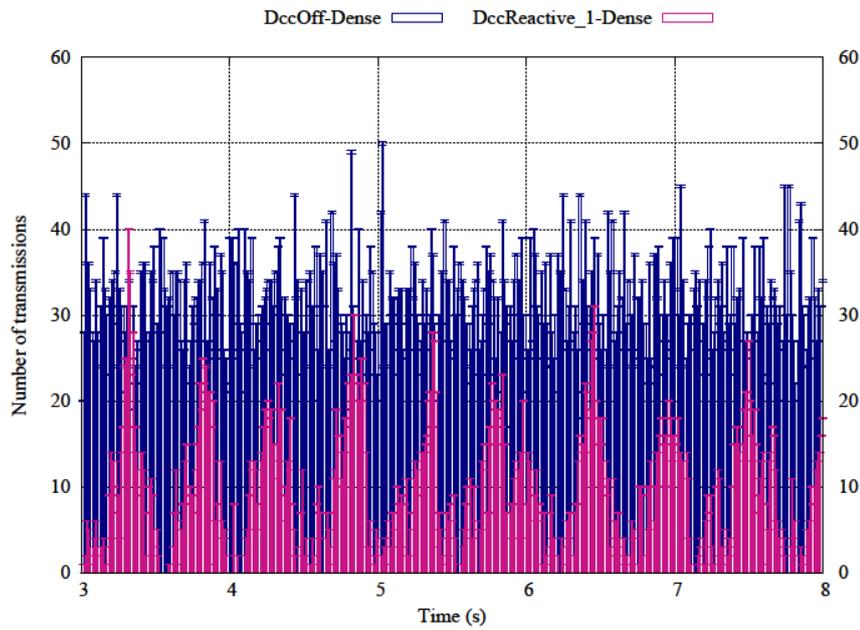

FIGURE 11: The distribution of the number of transmissions for DccReactive-1 and DCC-Off schemes (during a 5-second interval for dense scenario).

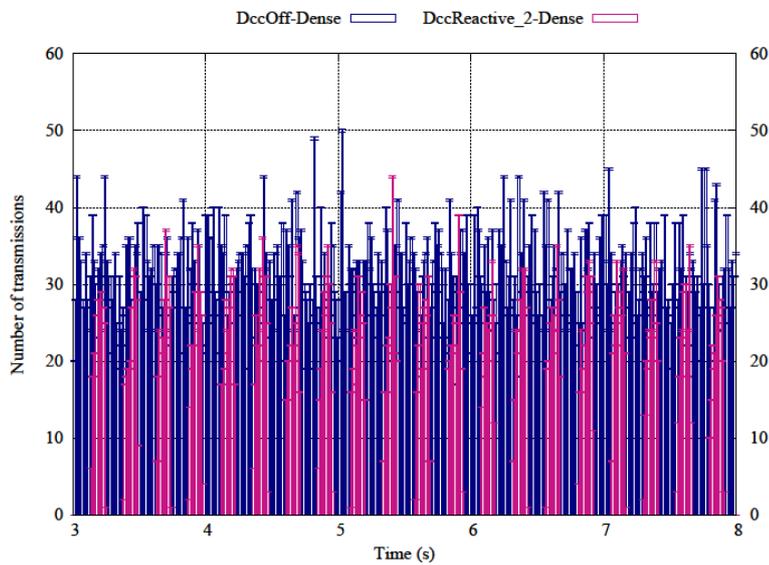

FIGURE 12: The distribution of the number of transmissions for DccReactive-2 and DCC-Off schemes (during a 5-second interval for dense scenario).

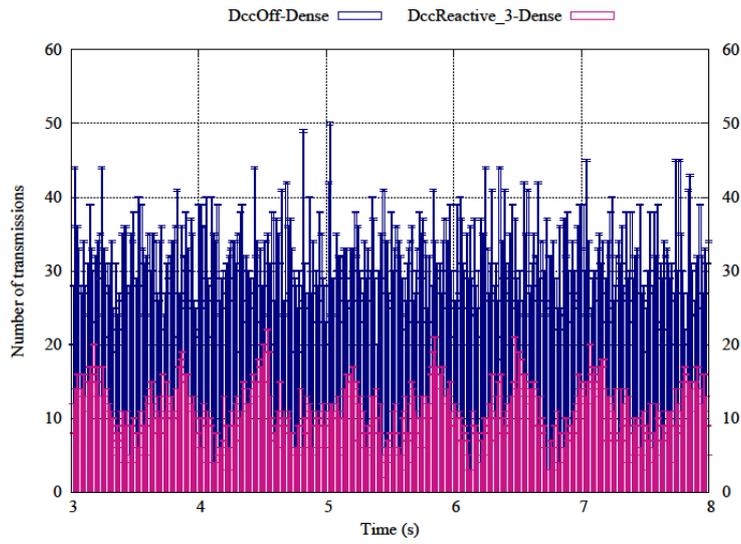

FIGURE 13: The distribution of the number of transmissions for DccReactive-3 and DCC-Off schemes (during a 5-second interval for dense scenario).

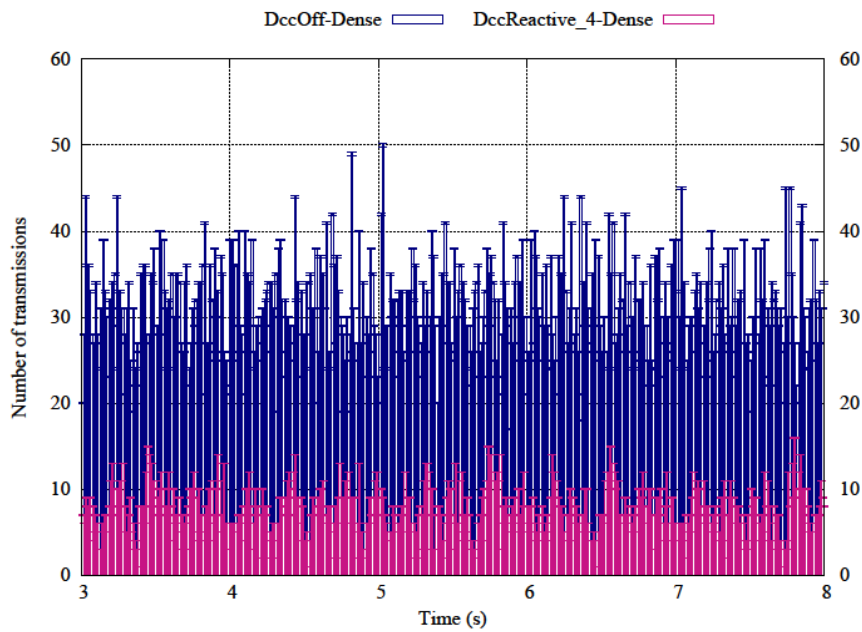

FIGURE 14: The distribution of the number of transmissions for DccReactive-4 and DCC-Off schemes (during a 5-second interval for dense scenario).

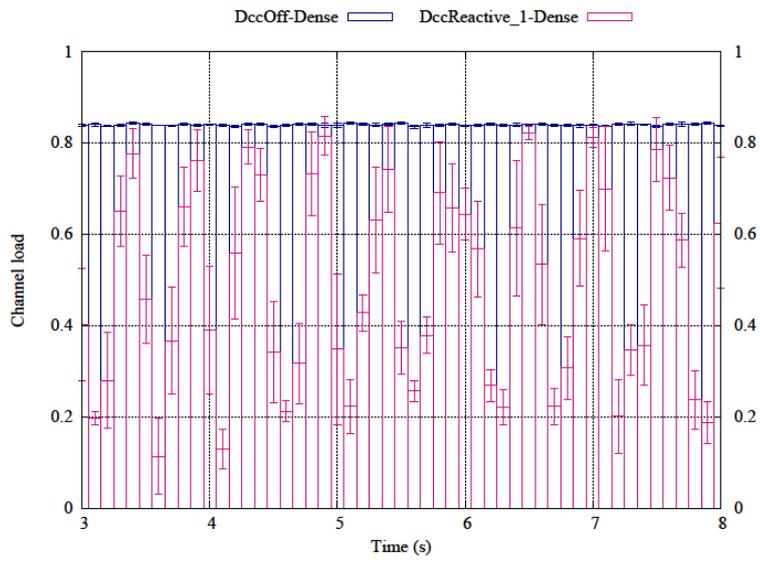

FIGURE 15: Average CBR during a 5-seconds of time interval for DccReactive-1 and DCC-OFF schemes.

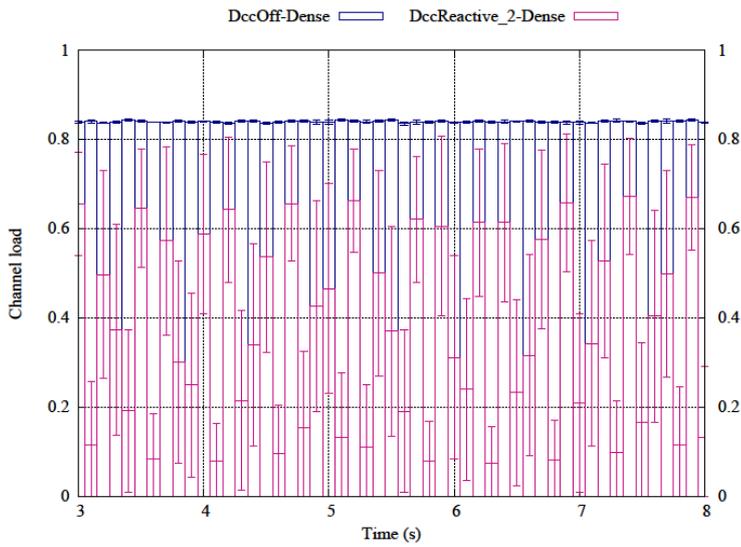

FIGURE 16: Average CBR during a 5-seconds of time interval for DccReactive-2 and DCC-OFF schemes.

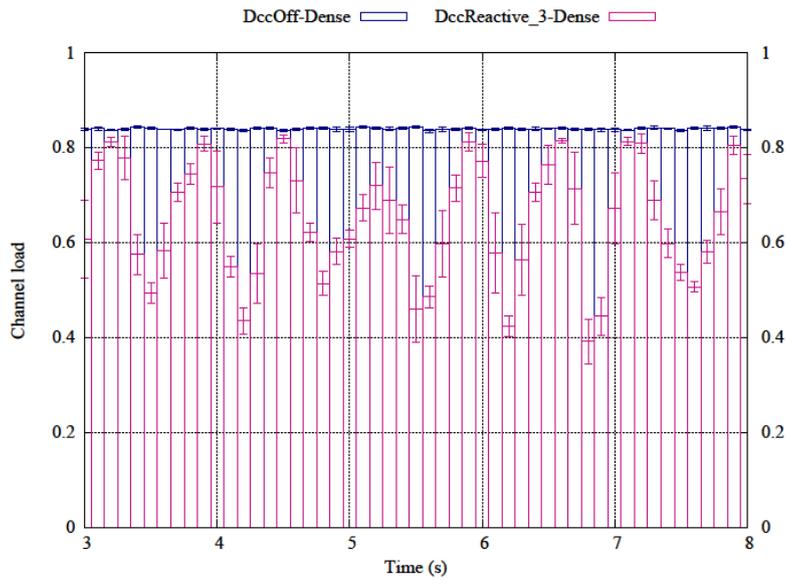

FIGURE 17: Average CBR during a 5-seconds of time interval for DccReactive-3 and DCC-OFF schemes.

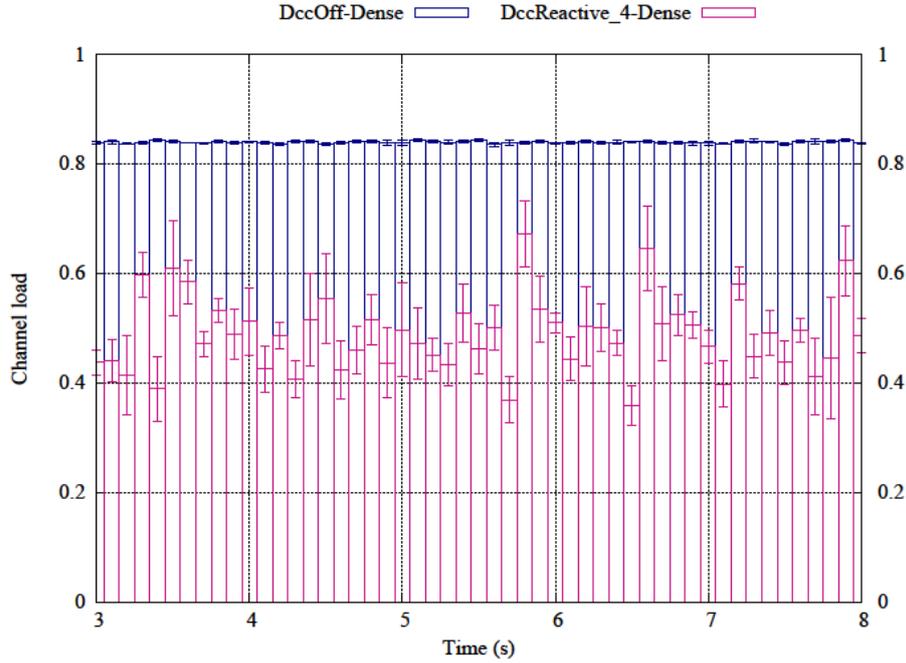

FIGURE 18: Average CBR during a 5-seconds of time interval for DccReactive-1 and DCC-OFF schemes.

### 4.4 Behavior of CAM Rate Control

Now we will have a closer look to the CAM generation behavior at an ITS-S. FIGURE 19- FIGURE 22 plot the setting and actual values of the CAM intervals as well as the measured CBR at a randomly selected ITS-S in the dense scenario. (Note that for visibility reason, the parameters are plotted only when the values change.) Similar to what it is seen in FIGURE 15-FIGURE 18, CBR fluctuates more for the synchronized mechanisms and less for the unsynchronized mechanisms. The setting value of the CAM interval tend to jump between the highest (460 ms) and the lowest (60 ms) values of the parameter table, Table 2, for the synchronized mechanisms (DccReactive-1 and -2). In the unsynchronized mechanisms, the CAM interval was set to large values (above 260 ms). Finally, CAMs tend to be transmitted at the intervals 1) equal to the setting intervals for DccReactive-1, 2) longer than the setting interval for DccReactive-2, 3) shorter and the equal to the setting interval in DccReactive-3, and 4) shorter or larger than the setting interval in Dcc-Reactive-4. Longer intervals than the setting values that observed in DccReactive-2 and -4 are conceivably due to the "Cancel-and-Go" behavior. Shorter intervals than the setting values that observed in DccReactive-3 and -4 are due to the "unsynchronized" behavior.

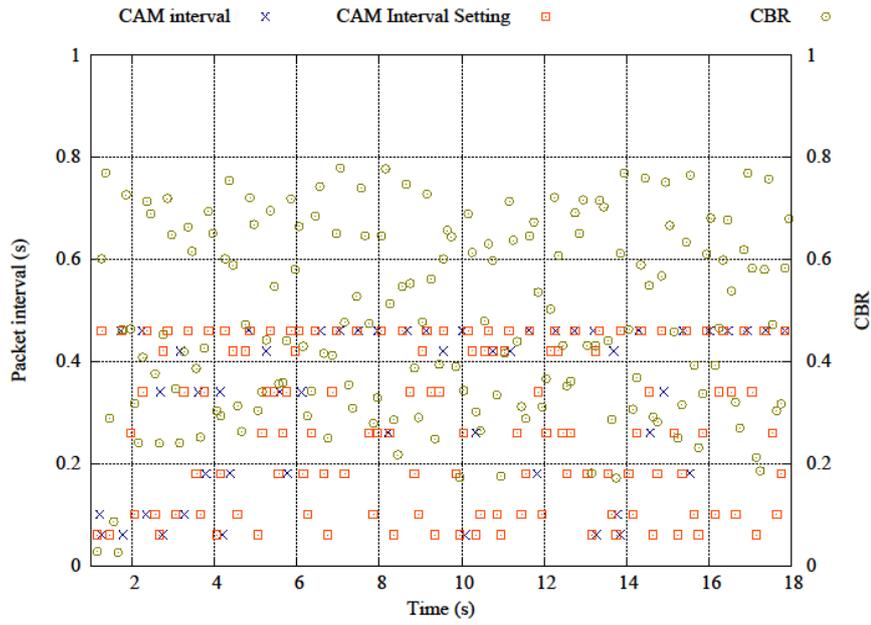

FIGURE 19: The setting (orange) and the actual (blues) values of the CAM interval and the measured CBR at a randomly selected ITS-S in the dense scenario for DccReactive-1.

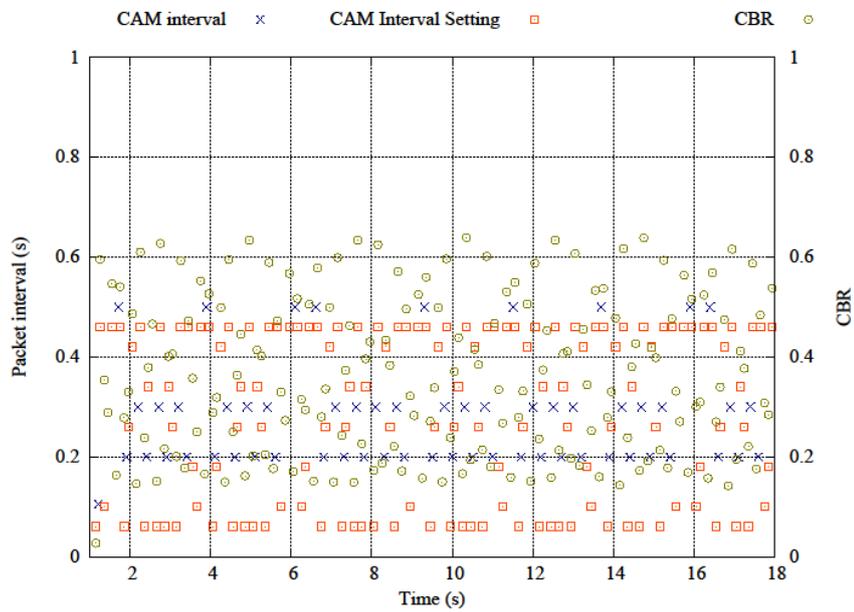

FIGURE 20: The setting (orange) and the actual (blues) values of the CAM interval and the measured CBR at a randomly selected ITS-S in the dense scenario for DccReactive-2.

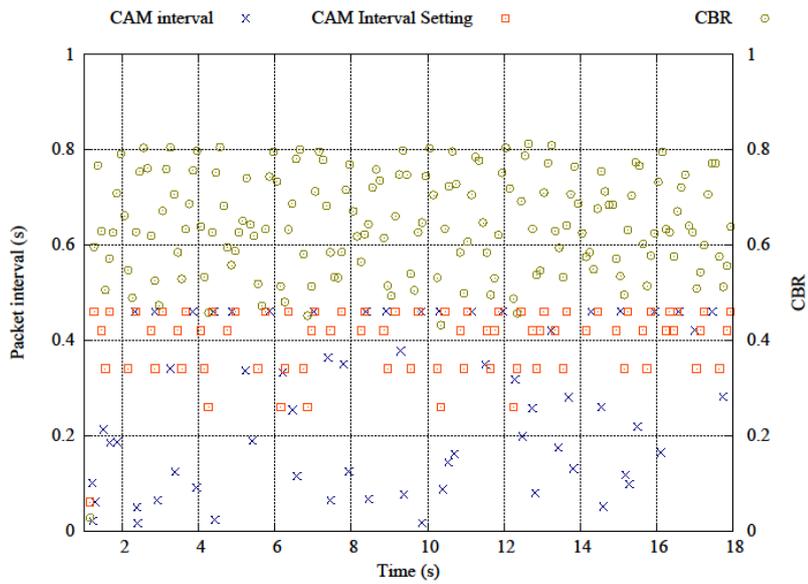

FIGURE 21: The setting (orange) and the actual (blues) values of the CAM interval and the measured CBR at a randomly selected ITS-S in the dense scenario for DccReactive-3.

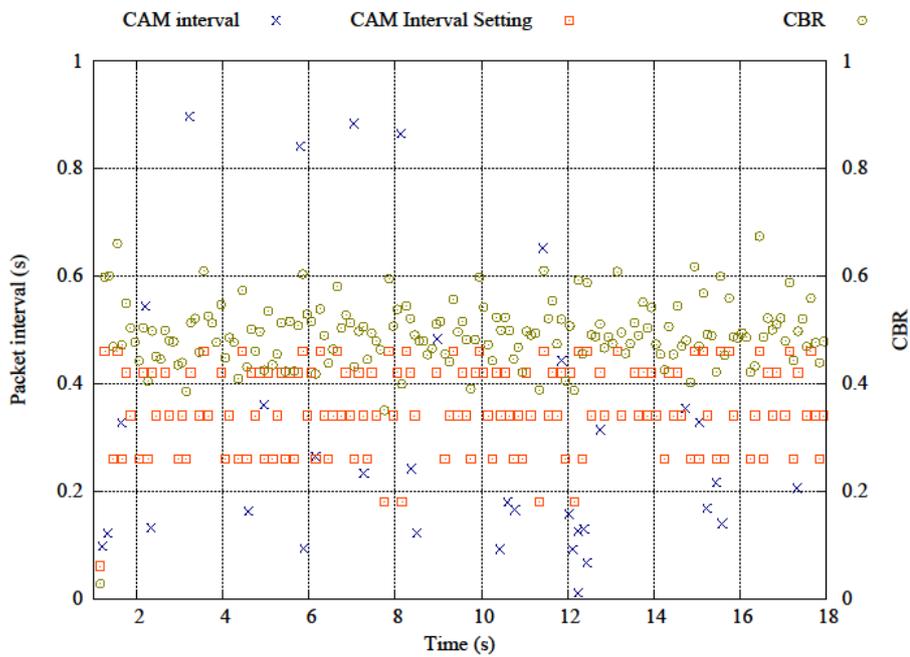

FIGURE 22: The setting (orange) and the actual (blues) values of the CAM interval and the measured CBR at a randomly selected ITS-S in the dense scenario for DccReactive-4.

The obvious observation can be made from FIGURE 11- FIGURE 22, is that the reactive DCC mechanisms in general use the wireless channel in an oscillated manner, and the oscillation is significant for the synchronized mechanisms (DccReactive-1 and DccReactive-2) especially for Cancel-and-Go & Synchronized scheme (DccReactive-2).

### 4.4 Summary

The table below lists the maximum positive and negative performance differences between the individual DccReactive and DccOff for the homogenous highway scenario. The table shows that in terms of PDR, DccReactive-4 (Cancel-and-Go & Unsynchronized) shows the best performance; but in terms of PIR the DccReactive-3 (Wait-and-Go & Unsynchronized) outperforms the other mechanisms. Since PDR is the ratio of the number of the received and the transmitted packets, a large PDR can be obtained by aggressively reducing the number of transmissions, i.e., CAM message generations. This is not always a good solution for safety applications. Indeed FIGURE 11- FIGURE 14 clearly show that the number of transmissions of DccReactive-4 is significantly smaller than that of DccReactive-3, explaining why DccReactive-4's PDR is better than that of DccReactive-3. On the other hand, PIR is the length of time during which the receiver node does not receive data from the transmitter node. For CAM packets, this obviously implies the time gap during which the receiver ITS-S does not have information about the transmitter ITS-S. Therefore PIR is one of the key parameters that can determine whether the V2X communications protocol can support such safety applications. To this reason, by paying more attention on PIR performances, and we conclude that DccReactive-3, Wait-and-Go & Unsynchronized, mechanism is the best approach among the four versions of DccReactive.

Table 5. Maximum performance improvement/deterioration of DccReactive schemes w.r.t DccOff.

| Algorithms | PDR difference ($PDR_{DccReactive} - PDR_{DccOff}$) | | PIR difference ($PIR_{DccReactive} - PIR_{DccOff}$) | |
|---|---|---|---|---|
| | Max improvement (Positive difference) [%] | Max deterioration (Negative difference) [%] | Max improvement [s] | Max deterioration [s] |
| DccReactive-1 | 44 | -2 | 0.22 | -0.43 |
| DccReactive-2 | 16 | -2 | 0 | -1.16 |
| DccReactive-3 | 68 | -1 | 0.68 | 0.22 |
| DccReactive-4 | 71 | -0.5 | 44 | 0.36 |

## 5.   Study on Channel Load Characterization

In this section, we study the impact of the weight factor a of the channel load defined in (1). The study is made based on the performance investigations of DccReactive-1 and -3 for homogeneous static highway scenario. Note that we omit the Cancel-and-Go mechanisms (DccReactive-2 and -4), because our results presented in Section 4 show that Cancel-and-Go mechanisms show degraded performance compared to the Wait-and-Go schemes. For simplicity, we now call DccReactive-1 as SyncDccReactive and DccReactive-3 as UnsyncDccReactive.

5.1 Simulation Results

FIGURE 23-FIGURE 26 plot the PIR performances of SyncDccReactive scheme for different density scenarios. Similarly, FIGURE 27-FIGURE 30 plot the PIR performances of UnsyncDccReactive schemes for different density scenarios. Each graph has several curves for different transmitter and receiver distance ranges. Specifically, the curve for "d" indicates the results obtained for transmitters and receivers that are at the distance [d-20, d+20] m from each other. The horizontal axis is a (in percentage), and hence we are interested in a, which provides the smallest (or relatively small) PIR value. As can be seen in FIGURE 23-FIGURE 26, no particular value of that provides satisfying performance can be found for SyncDccReactive for all the scenarios and distance ranges. We can say however, a =1, where the algorithm considers only the last CBR value, tends to lead to poorer performances for SynDccReactive. In contrast, for UnsyncDccReactive (see FIGURE 27-FIGURE 30), a =1 provides the best performance almost all the scenarios and distance ranges.

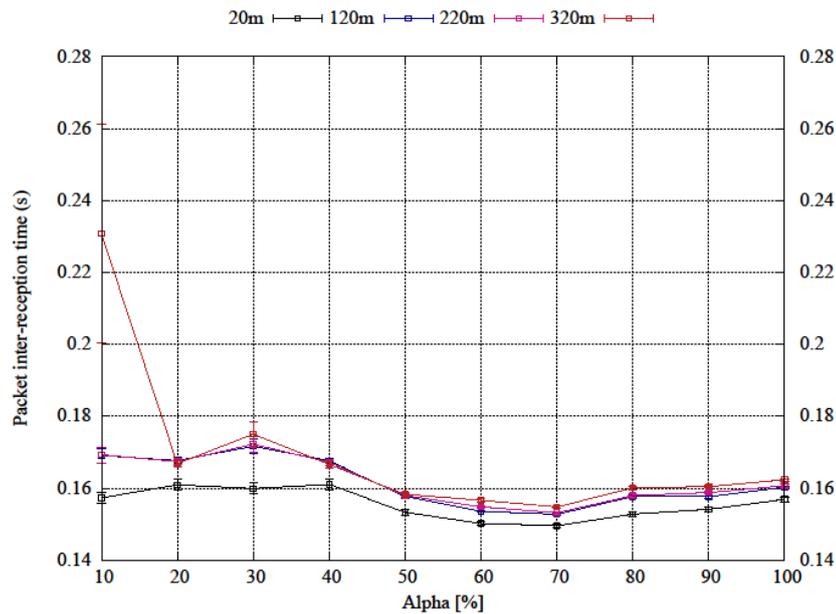

FIGURE 23: PIR of SyncDccReactive for different values of the weight factor (α) in the sparse scenario.

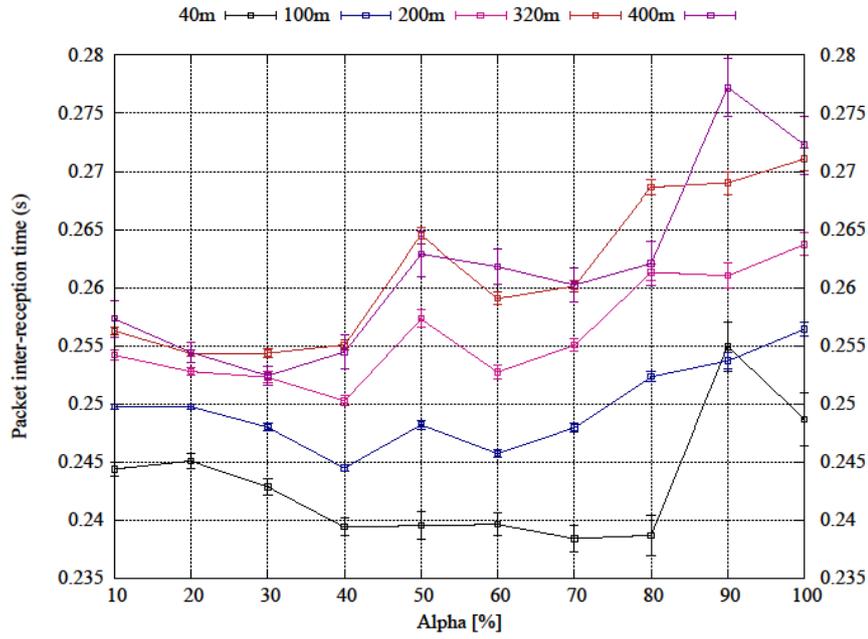

FIGURE 24: PIR of SyncDccReactive for different values of the weight factor (α) in the medium scenario.

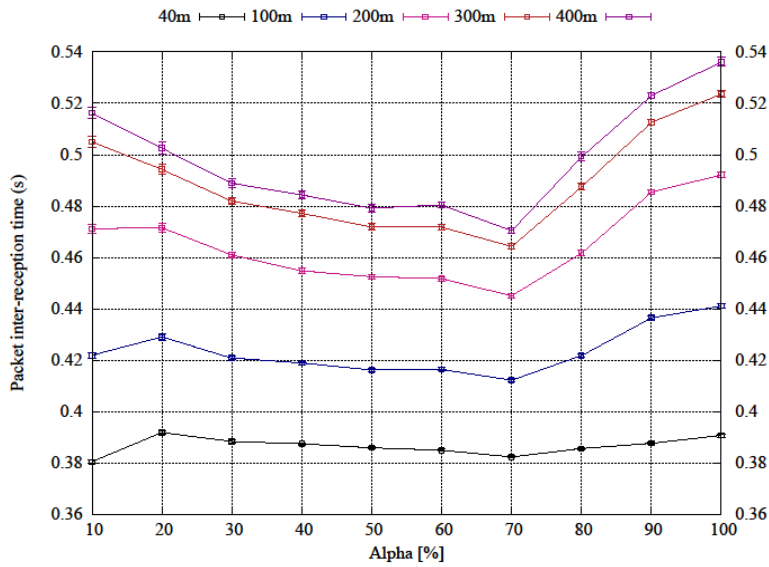

FIGURE 25: PIR of SyncDccReactive for different values of the weight factor (α) in the dense scenario.

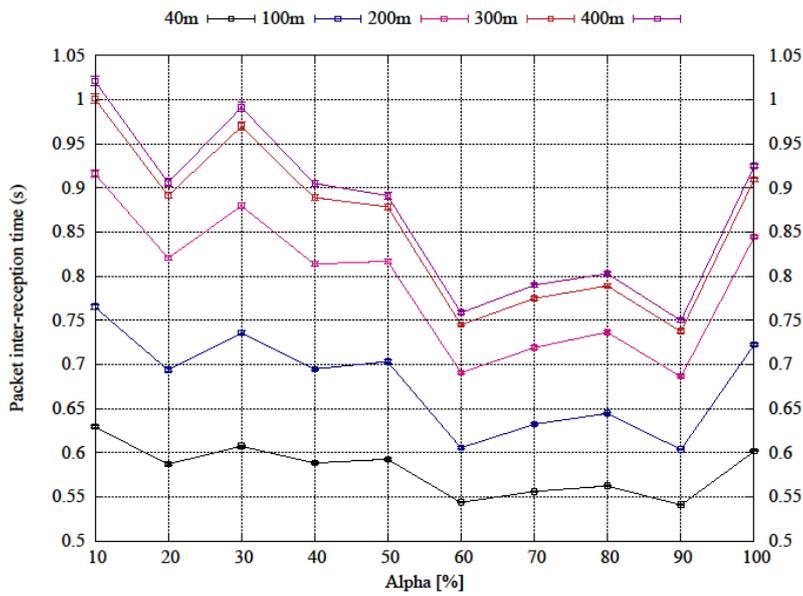

FIGURE 26: PIR of SyncDccReactive for different values of the weight factor (α) in the extreme scenario.

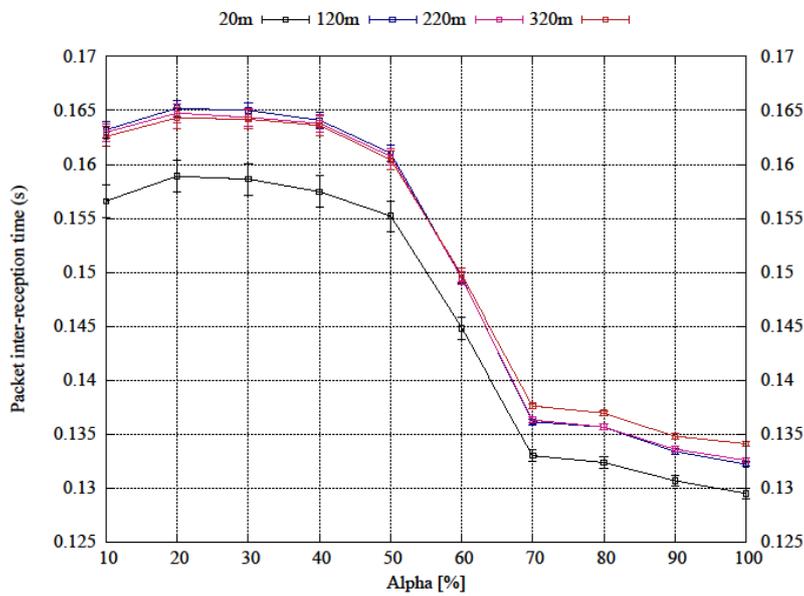

FIGURE 27: PIR of UnsyncDccReactive for different values of the weight factor (α) in the sparse scenario.

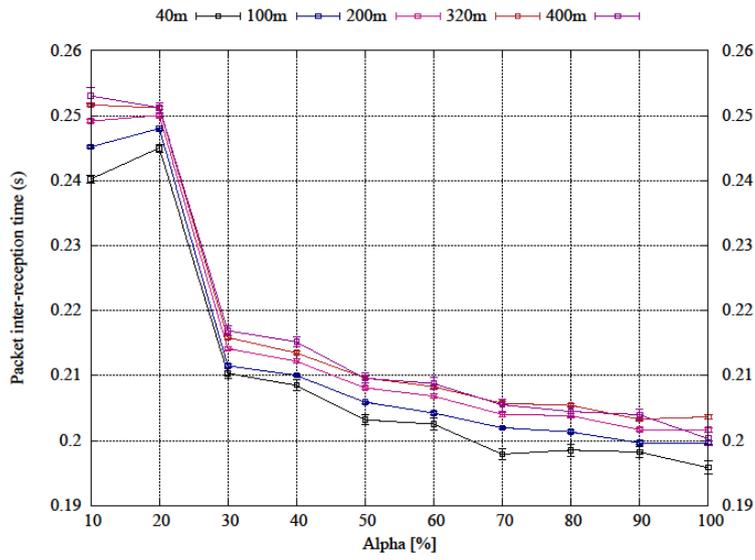

FIGURE 28: PIR of UnsyncDccReactive for different values of the weight factor (α) in the medium scenario.

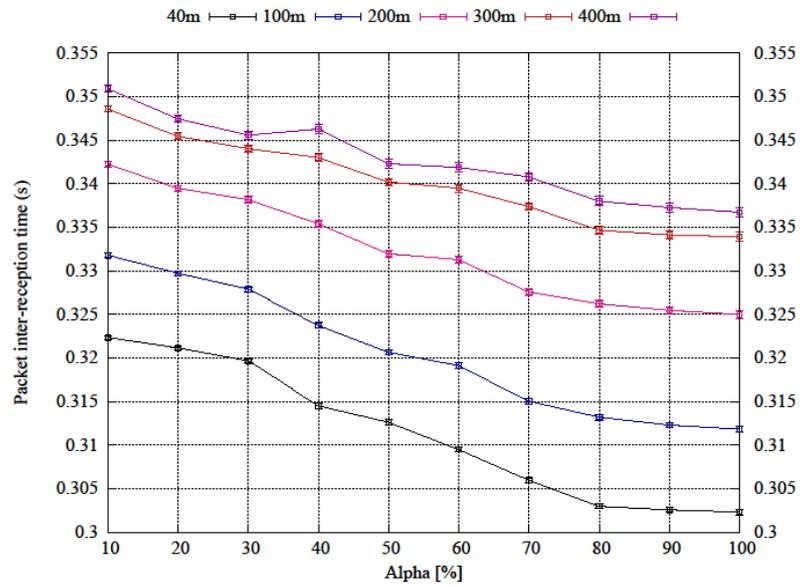

FIGURE 29: PIR of UnsyncDccReactive for different values of the weight factor (α) in the dense scenario.

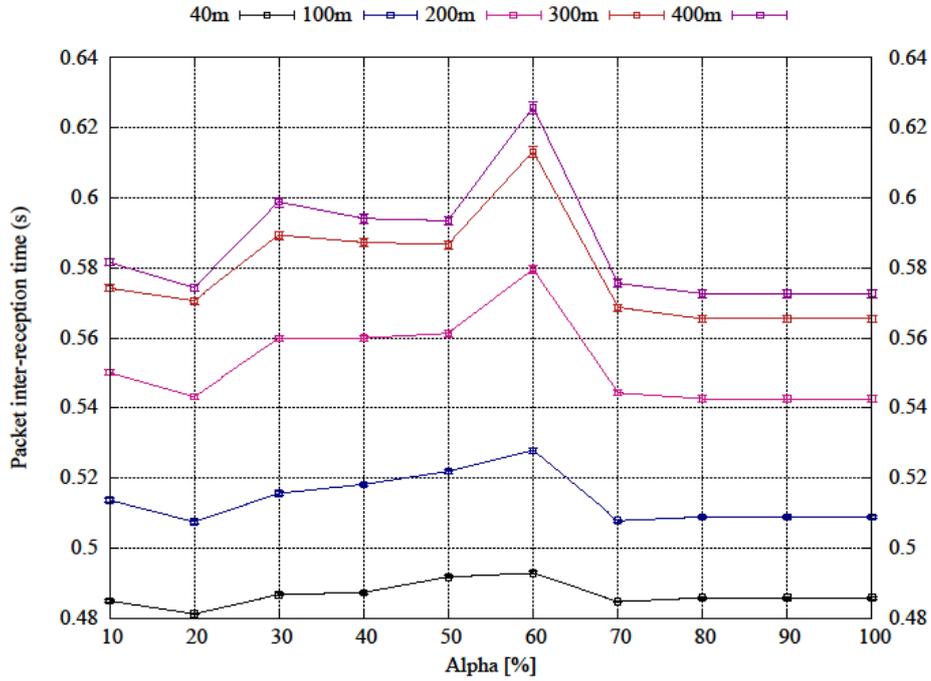

FIGURE 30: PIR of UnsyncDccReactive for different values of the weight factor (α) in the extreme scenario.

## 5.2 Summary

Table 6 and Table 7 list the weight factor, α (see (1)), which corresponds to the shortest PIR for SyncDccReactive (DccReactive_1) and UnsyncDccReactive (DccReactive_3) mechanisms. The difference of PIR between the minimum value and that for α =1 is indicated for the cases, where α is not 1. Table 6 clearly shows that for the synchronized DCC system, the best PIR performances never achieved when α =1; the difference between the minimum PIR and that for α =1 is over 10 milliseconds. This implies that it is difficult to characterize channel load by only the current CBR. In contrast, as Table 7 shows that when the system is unsynchronized, the best PIR is achieved when α =1, indicating that channel load can be characterized by only the current CBR if the system is unsynchronized. Note that α =0.2 provides the shortest PIR for small transmitter and receiver distances in the extreme scenario for UnsyncDccReactive, the PIR difference between the minimum value that that for α =1 is very small (4ms).

Table 6. The weight factor, α, which corresponds to the shortest PIR for SyncDccReactive.

| Scenario | Transmitter and Receiver distance range [m] | | | | |
|---|---|---|---|---|---|
| | 40 m | 100 | 200 | 300 | 400 |
| Medium | 0.7 (10ms shorter than $PIR(\alpha = 1)$) | 0.4 (12ms shorter than $PIR(\alpha = 1)$) | 0.4 (13ms shorter than $PIR(\alpha = 1)$) | 0.2 (16ms shorter than $PIR(\alpha = 1)$) | 0.3 (20ms shorter than $PIR(\alpha = 1)$) |
| Dense | 0.2 (10ms shorter than $PIR(\alpha = 1)$) | 0.1 (29ms shorter than $PIR(\alpha = 1)$) | 0.7 (47ms shorter than $PIR(\alpha = 1)$) | 0.7 (59ms shorter than $PIR(\alpha = 1)$) | 0.7 (66ms shorter than $PIR(\alpha = 1)$) |
| Extreme | 0.1 (61ms shorter than $PIR(\alpha = 1)$) | 0.9 (12ms shorter than $PIR(\alpha = 1)$) | 0.9 (16ms shorter than $PIR(\alpha = 1)$) | 0.9 (17ms shorter than $PIR(\alpha = 1)$) | 0.9 (17ms shorter than $PIR(\alpha = 1)$) |

Table 7. The weight factor, α, which corresponds to the shortest PIR for UnsyncDccReactive

| Scenario | Transmitter and Receiver distance range [m] | | | | |
|---|---|---|---|---|---|
| | 40 | 100 | 200 | 300 | 400 |
| Medium | 1 | 1 | 1 | 1 | 1 |
| Dense | 1 | 1 | 1 | 1 | 1 |
| Extreme | 0.2 (4 ms shorter than that of $\alpha = 1$) | 0.2 (4 ms shorter than that of $\alpha = 1$) | 1 | 1 | 1 |

## 6. Study on Non-Identical Sensing Capabilities

In this section, we study the impact of non-identical sensing capabilities. Based on the previous work, this study targets DccReactive-3 (Wait-and-Go & Unsynchronized) and DccOff mechanisms.

### 6.1 Simulation Results

Figure 31 - Figure 32 compare PDR of DccReactive and DccOff mechanisms for the cases where ITS-Ss have identical and non-identical sensing capabilities, respectively. The horizontal axis is the distance between the transmitters and the receivers. As can be seen in Figure 31, if the ITS-Ss have identical capabilities, the average PDR is stable when the distance between the transmitter and receiver is below 420 meters for both DccOff and DccReactive mechanisms. On the other hand, when the system consists of non-identical ITS-Ss (Figure 32), the stable distance is up to 250 meters for the sparse scenario and shorter for the medium and high density scenarios. Figure 33 - Figure 34 compare PIR of DccReactive and DccOff mechanisms for the cases where ITS-Ss have identical and non-identical sensing capabilities, respectively. A similar observation as for the case of PDR can be made for PIR, which can reach 1.6 seconds for DccOff for non-identical sensing capability.

In Figure 35 - Figure 36, Jain's fairness index is calculated targeting the number of transmissions for DccReactive and DccOff mechanisms for the cases where ITS-Ss have identical and non-identical sensing capabilities, respectively. The horizontal axis shows the density classes: 100m, 45m, 20m, and 10m represent the sparse, medium, dense, and extreme classes. In the case of the identical ITS-Ss, the fairness index is 100% for DccOff regardless of density class, and it is slightly lower for DccReactive. In contrast, in the case of non-identical ITS-Ss, the fairness index degrades, and performance degradation is significant for DccReactive.

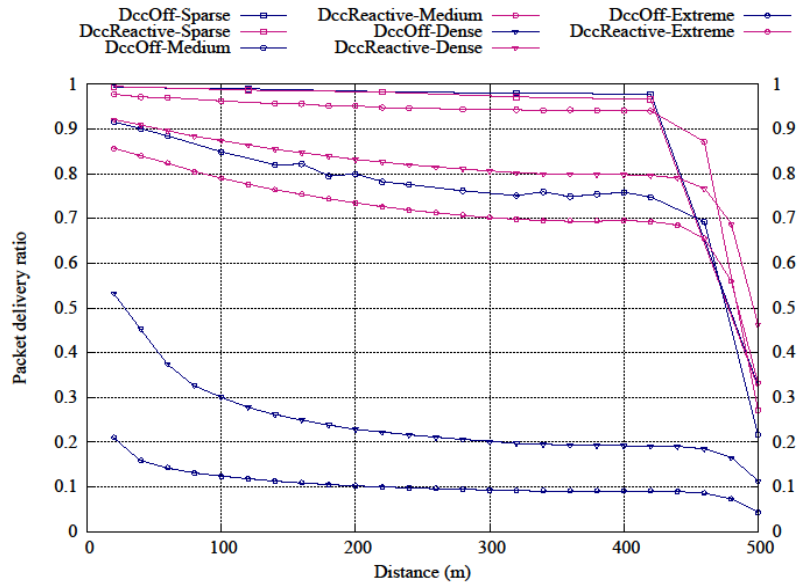

FIGURE 31: Comparison of PDR of DccReactive and DccOff schemes for the systems with ITS-Ss with identical receiver capabilities.

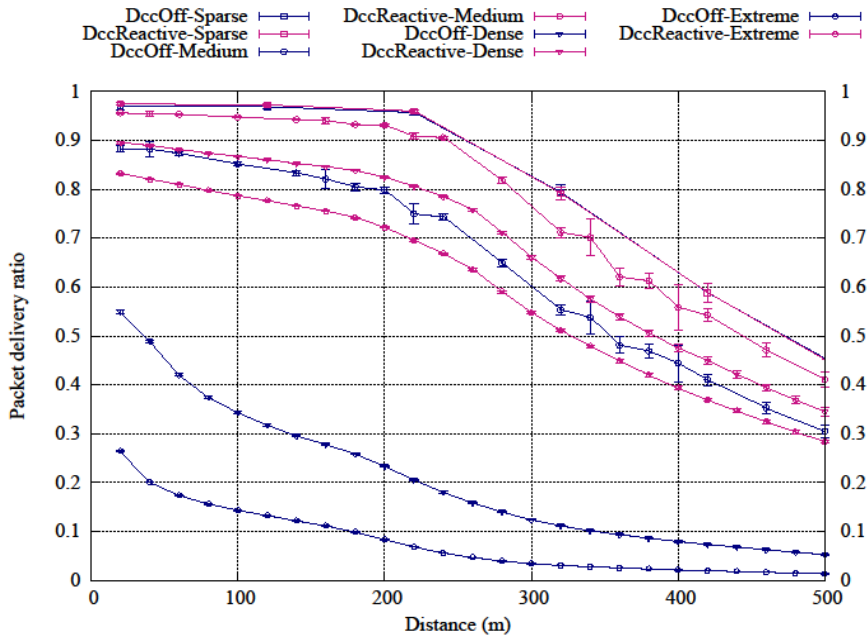

FIGURE 32: Comparison of PDR of DccReactive and DccOff schemes for the systems with ITS-Ss with non-identical receiver capabilities.

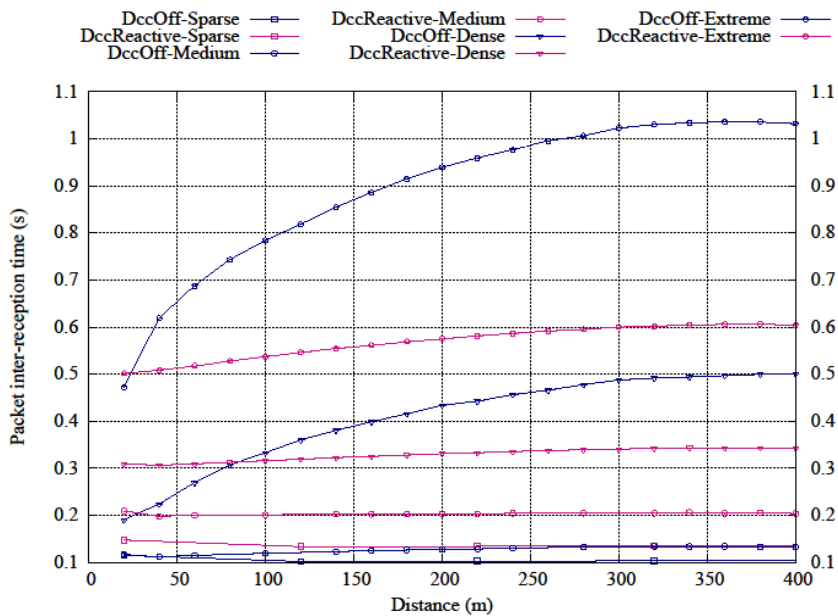

FIGURE 33: Comparison of PIR of DccReactive and DccOff schemes for the systems with ITS-Ss with identical receiver capabilities.

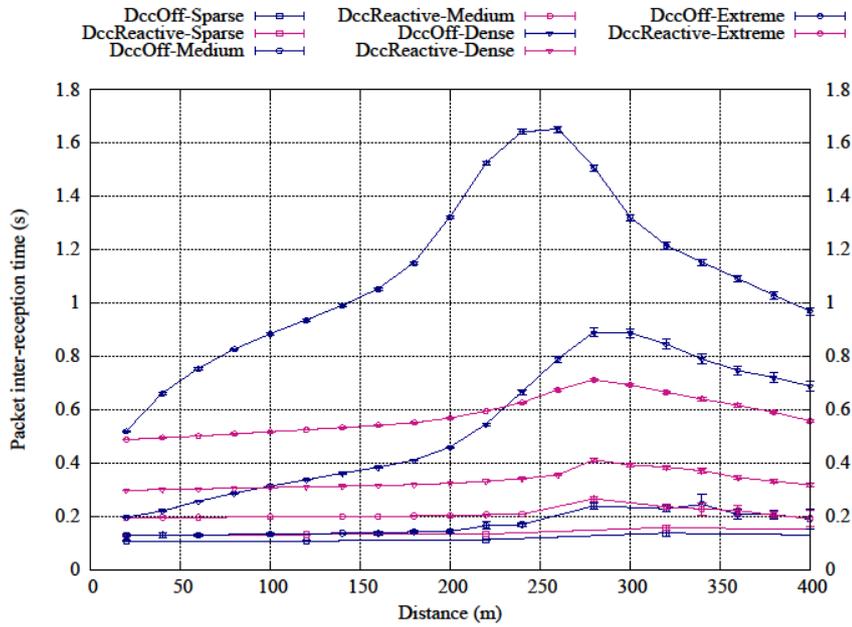

FIGURE 34: Comparison of PIR of DccReactive and DccOff schemes for the systems with ITS-Ss with non-identical receiver capabilities.

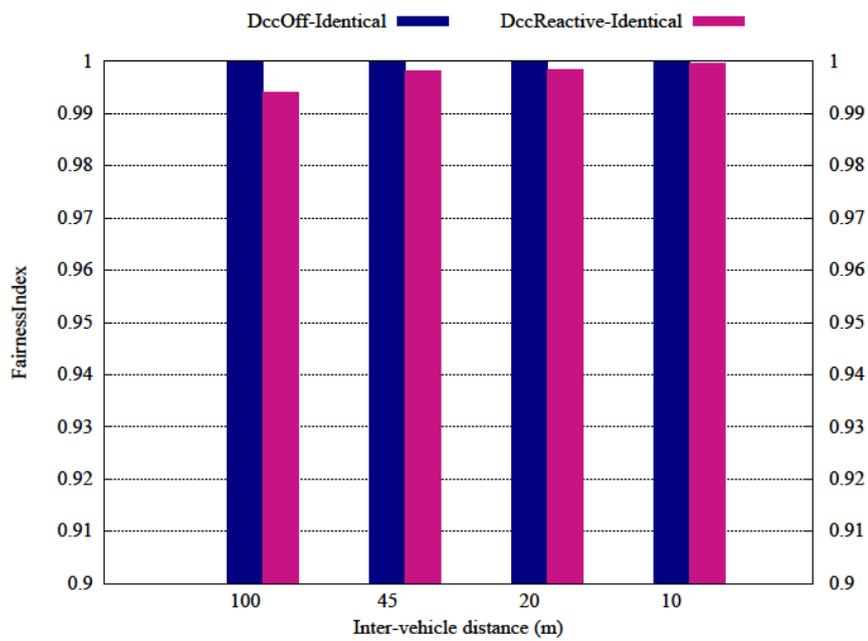

FIGURE 35: Comparison of fairness index for DccReactive and DccOff schemes for the systems with ITS-Ss with identical receiver capabilities.

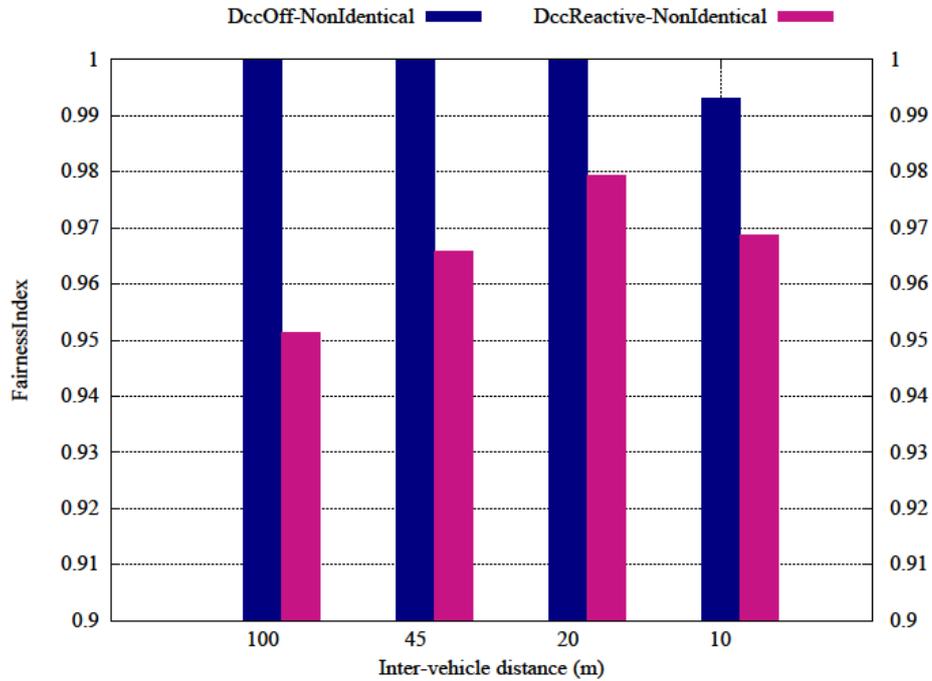

FIGURE 36: Comparison of fairness index for DccReactive and DccOff schemes for the systems with ITS-Ss with identical receiver capabilities.

## 6.2 Summary

The results clearly show that when the ITS-Ss have non-identical sensing capabilities, in general, the average communication range tends to be reduced for both DccOff and DccReactive schemes. The unfair sensing capabilities result in unfair transmission behaviors. The unfairness issue is more significant for DccReactive than for DccOff.

## Conclusion

In this work, we studied the following issues targeting reactive dynamic DCC algorithm.
- Synchronization
- Channel load characterization
- Non-identical receiver capability

Following conclusions were drawn:
- It is very important to provide a solution to avoid synchronized DCC behaviour among ITS-Ss. If a careful attention is given on this issue, the simple reactive DCC algorithm can perform better than DccOff. In the case of rate adaptation, introducing a random rate seems to be a good solution.
- If the road traffic is sparse, the reactive DCC algorithm tends to show poorer performance than DccOff.
- Cancelling timer for the CAM generator seems to be not necessary.
- If the system is unsynchronized, it seems that only the current CBR can be a good indicator of the channel load. However, if the system is synchronized, it is necessary to pay attention on CBR for longer interval.
- If the system consists of ITS-Ss with heterogeneous channel sensing capability, non-negligible negative impact can be expected in terms of communications range and fairness.
- The fairness issue caused by non-identical sensing capabilities is more significant for DCC-enabled system.

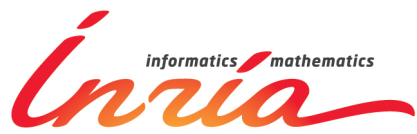